\documentclass[11pt]{article}
\usepackage{moriond,epsfig}

\def\be{\begin{equation}}
\def\ee{\end{equation}}
\def\bea{\begin{eqnarray}}
\def\eea{\end{eqnarray}}

\def\lesssim{\mathrel{\hbox{\rlap{\hbox{\lower4pt\hbox{$\sim$}}}\hbox{$<$}}}}
\def\gtrsim{\mathrel{\hbox{\rlap{\hbox{\lower4pt\hbox{$\sim$}}}\hbox{$>$}}}}

\begin{document}
\title{Observation of Gamma-Ray Bursts with INTEGRAL}

\author{D. G\"{o}tz$^{1,2}$ and S. Mereghetti$^{1}$}

\address{$^{1}$Istituto di Astrofisica Spaziale e Fisica Cosmica - Sezione di Milano ``G.Occhialini'' - CNR, Italy \\
$^{2}$Dipartimento di Fisica -- Universit\`a degli Studi di Milano-Bicocca, Italy}

\maketitle\abstracts{
We expect that at least one Gamma-Ray Burst per month
will be detected in the field of view of INTEGRAL instruments
and localized with an accuracy of a few arcminutes.\\
IBIS (the  Imager on board INTEGRAL) will be the most sensitive instrument in the soft
$\gamma$-ray band   during the next few years. This
will allow INTEGRAL to detect the faintest (and hence
the most distant)  $\gamma$-ray bursts, thus surely contributing
to the study of the physics of GRBs and
to the comprehension of the early stages of the Universe.\\
GRBs will be
detected and localized in (near) real time with the IBAS (INTEGRAL
Burst Alert System) software, and their coordinates will be immediately
distributed to the scientific community, allowing
for prompt follow-up observations at other wavelengths.
The flexibility of the IBAS software, running on ground at the ISDC,
will allow us to possibly reveal new
classes of bursts,  such as ultra-short bursts ($\sim$ ms) or
very long (t $>$ 50 s) but slowly rising bursts, that could not be detected
by previous experiments.
}

\section{Introduction}

The origin and nature of Gamma-Ray Bursts (GRBs) have been a puzzling mystery
since their discovery in the late 60's \cite{klebesadel}.
This was mainly due to the fact that
these short and unpredictable  events had been observed exclusively at $\gamma$-ray
energies and with non-imaging instruments.
A great advance  came after the launch of CGRO in 1991: BATSE registered a great
number ($\sim$3000) of GRBs isotropically distributed over the entire
sky, but with a non-euclidean LogN-LogS slope, suggesting a cosmological origin.
The BATSE results contributed to develop
a {\it standard} model for the emission mechanisms of the prompt and the delayed emission:
the so called internal \cite{rees94}
and external \cite{rees92} fireball shock model, involving
a relativistically expanding fireball with bulk Lorentz factors of the order
of $\sim$ 100. This model is independent of the nature and details of
the GRB progenitors. In fact, despite substantial efforts no certain evidence
on the progenitors has been found yet.
The cosmological nature of GRBs was confirmed after the discovery of the first
(predicted) X--ray afterglow in 1997 (GRB970228) by {\it Beppo}SAX \cite{costa},
which allowed the first successful follow-up at optical wavelengths \cite{vanpa}
and hence its red-shift determination \cite{djorgovski} (z=0.695).
However, the debate
on the progenitors of these sources is still open. While for the long GRBs (t $>$ 2 s)
the models involving the core collapse of a very massive star seem to be appropriate
(e.g. \cite{woosley}),
very little is known on the short GRBs (t $<$ 2 s), for  which no counterparts at
longer wavelength have been observed yet.
In their case, the model of the coalescence between compact
objects (NS-NS, NS-BH) \cite{meszaros} cannot be ruled out.\\
In this framework a clear requirement of the scientific community is
the rapid localization of the prompt emission of GRBs followed by an efficient
distribution of the derived coordinates.
The INTEGRAL satellite, although not specifically designed as a GRB oriented
mission, can greatly contribute to these studies. Its main imaging instrument,
IBIS \cite{ubertini}, is very
sensitive in the energy range from $\sim$20 keV to a few MeV
and, coupled with the IBAS software (see $\S$ \ref{IBAS}) running on ground, it will
provide in near real time the positions of the GRBs detected in its large
field of view. This can be particularly useful in the case of short
GRBs (see $\S$ \ref{perf}) to confirm or deny the presence of an afterglow at other wavelengths.
In addition, being more sensitive than previous instruments
it will be able to detect and localize even faint GRBs at high  red-shifts (z $>$ 10),
which could be associated with Population III stars, allowing us to
investigate the early Universe.

\section{The INTEGRAL Burst Alert System (IBAS)}
\label{IBAS}
The INTEGRAL Burst Alert System (IBAS) is a software system which provides the
(near) real time detection and localization of GRBs. The software is run on ground at the
INTEGRAL Science Data Centre (ISDC) \cite{courvoisier} and the derived positions are immediately distributed
to the scientific community to allow for follow up observations at other wavelengths.
Thanks to the properties of the ISGRI detector (IBIS upper layer covering the
lowest energy range) the accuracy of the positions is of
\begin{figure}[ht]
\centerline{\epsfig{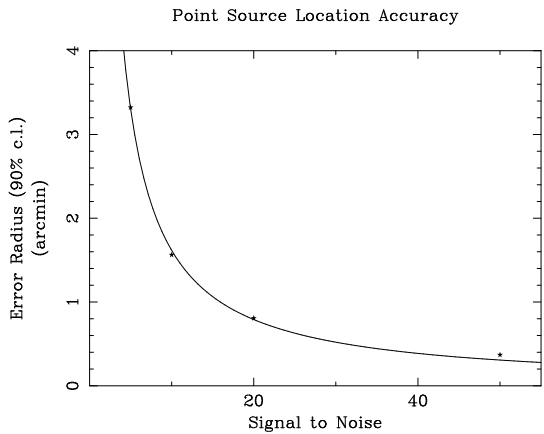}}
\vspace{10pt}
\caption{ISGRI point source location accuracy. The line represents a theoretical
calculation and the points simulated data (Mereghetti et al. 2001).}
\label{sla}
\end{figure}
the order of a few arcminutes or better (see figure \ref{sla}).\\
Thanks to the fact that the software is run on ground, there is great flexibility in the
choice of parameters and in the optimization of the triggering algorithms.
For ISGRI two different methods  have been implemented to search for GRBs: one,
based only on imaging,  looks for the appearance of new sources by comparing the
instantaneous image of the sky with those obtained  previously.
The other one looks for statistically significant excesses in the overall count rate
on various time scales and uses the imaging only as a confirmation to distinguish
real sources from background variations or other instrumental effects.
The latter algorithm is particularly
efficient for short burst, while the former can successfully be used in the case
of slowly rising bursts (see $\S$ \ref{perf}).
Multiple
instances of both  algorithms  can run simultaneously with different parameters, such as
energy channels interval, integration time scales, etc.
The IBAS software has also the capability to automatically reconfigure the
INTEGRAL Optical Monitoring Camera (OMC), by sending an appropriate telecommand,
 when a GRB is located within its field of view (5$^{\circ}\times$
5$^{\circ}$, but only a limited number of predefined small windows are observed).
\subsection{IBAS Performances for very short and long GRBs}
\label{perf}
\begin{figure}[ht!]
\centerline{\epsfig{file=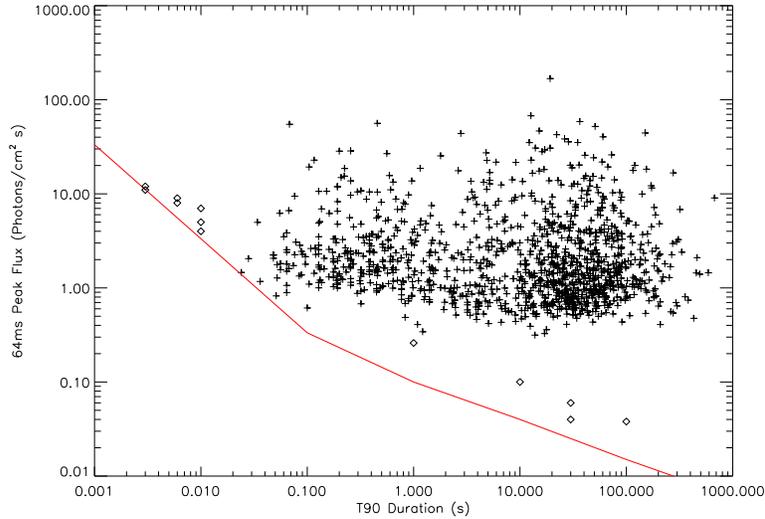,height=3.9in,angle=90}}
\caption{Crosses: BATSE on-flight triggers. Diamonds: IBAS ``successful'' simulations
(i.e. the GRB was detected and localized correctly)}
\label{sens}
\end{figure}

\begin{figure}[ht]
\begin{minipage}{2.5in}
\hspace*{-1.3cm}
\epsfig{figure=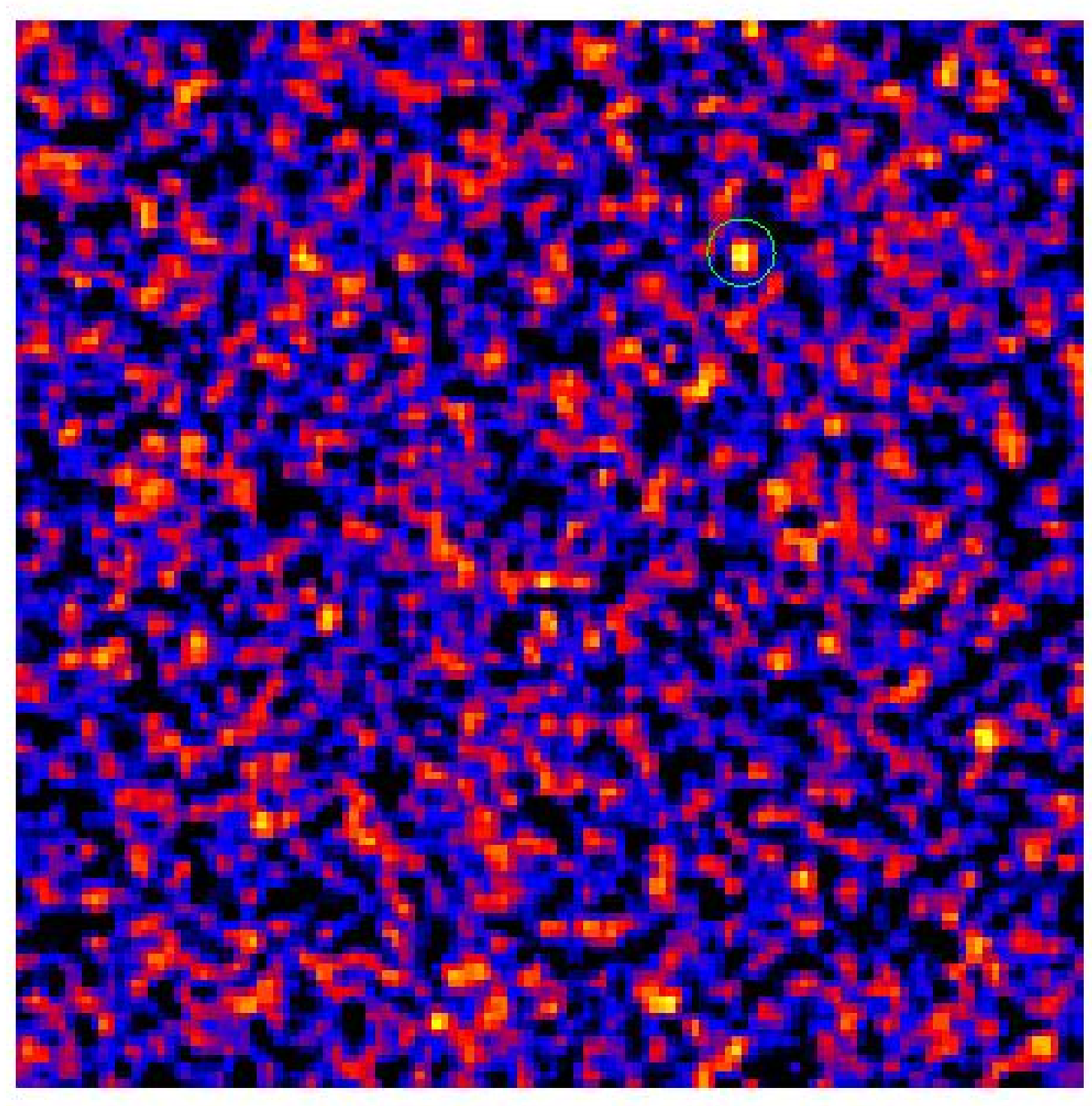 , height=3.3in,  angle=-90}
\caption{ISGRI image of a 10 ms burst generated with IBAS}
\label{10ms}
\end{minipage}
\hfill
\begin{minipage}{3.4in}
\hspace*{-0.6cm}
\epsfig{figure=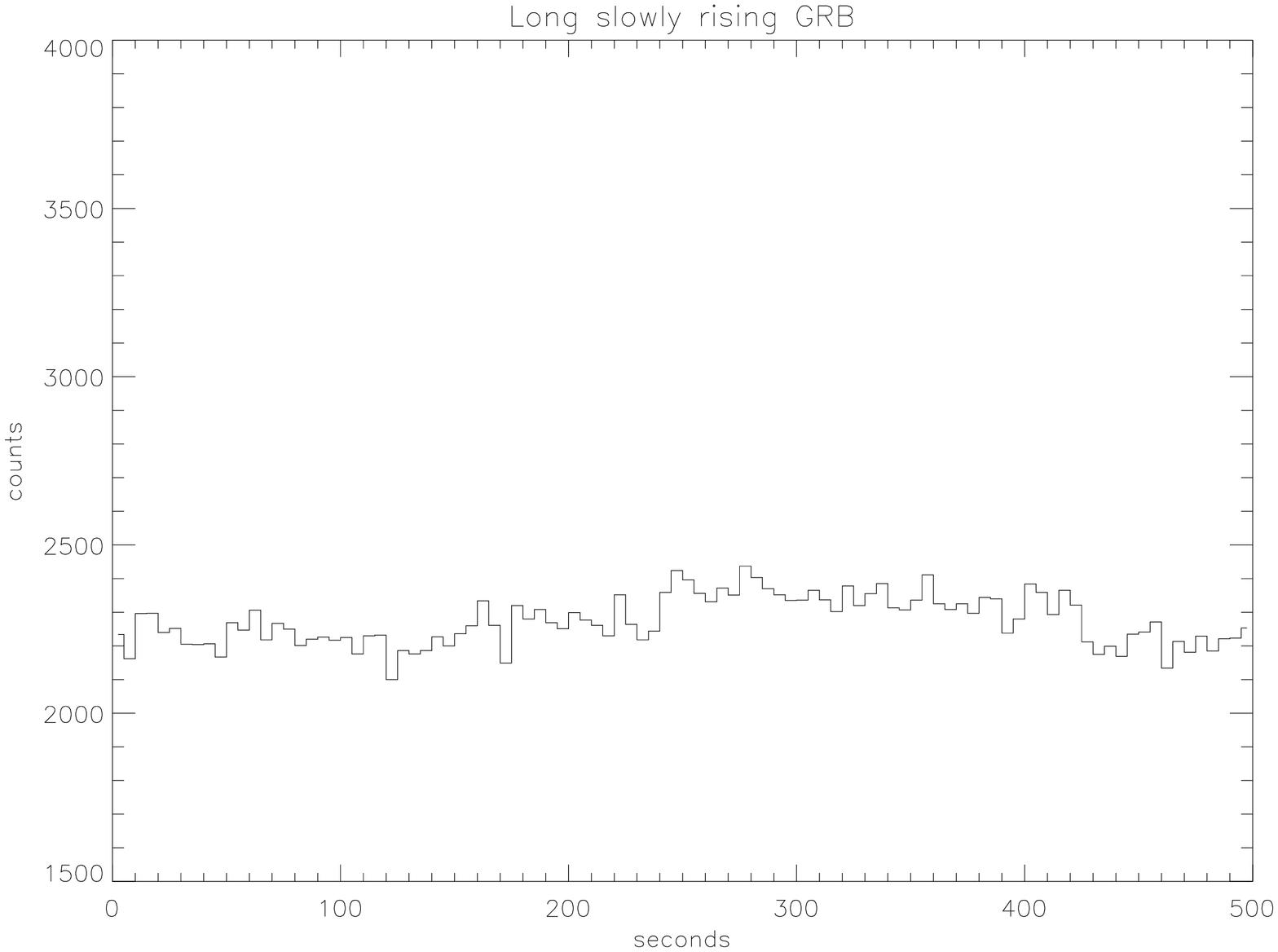,width=2.2in,height=3.8 in,angle=90}
\caption{IBAS light curve (15-300 keV, time bin=5 s) of a long but slowly rising burst. The burst begins at t=200 s and lasts for about 150 s.}
\label{long}
\end{minipage}
\end{figure}

We expect to detect and localize $\sim$1-2 bursts per month in the IBIS field of
view \cite{mereghetti} (29$^{\circ}$ Full Width Zero Response).
This estimate is based on the extrapolation of the
log N-log P distribution obtained with the BATSE on-flight triggers \cite{paciesas}.
IBIS will be the
most sensitive instrument in the soft $\gamma$-ray range during the next years
(note that Swift \cite{hurley},
which is based on a similar detector technology, has a greater effective area
(factor $\sim$2), but a higher
background due to the cosmic diffuse radiation since it has a larger field of view
(factor $\sim$9)).
As a result of the better sensitivity, we expect to extend the lower end of the BATSE
log N-log P distribution (detecting the faintest and hence the most distant burst ever),
to possibly discover a population of ultra-short bursts ($\sim$ ms) and to detect very long GRBs
(t $\gtrsim$ minutes) even if they have a  slowly rising time profile.\\
We have performed several simulations to show how the parameter space of detected
GRBs can be extended. This is shown in Figure \ref{sens}, where the IBAS sensitivity,
as estimated from theoretical calculations, and the results of the simulations are
plotted over the  Peak Flux vs. T90 duration of all the GRBs contained in the
BATSE 4B catalog \cite{paciesas}.
Figure \ref{10ms} shows the reconstructed 15-300 keV image of one of the
very short  bursts.
The light curve  of a very long and slowly rising GRB, that has been successfully detected
in one of our simulations, is shown in figure \ref{long}.
Such kind of GRBs, if they exist, are difficult or impossible to detect with
non-imaging instruments often operating in conditions of variable background, while
the IBAS imaging algorithm, when used with the appropriate parameters,
can detect them.
\subsection{X--ray Flashes}
\begin{figure}[ht!]
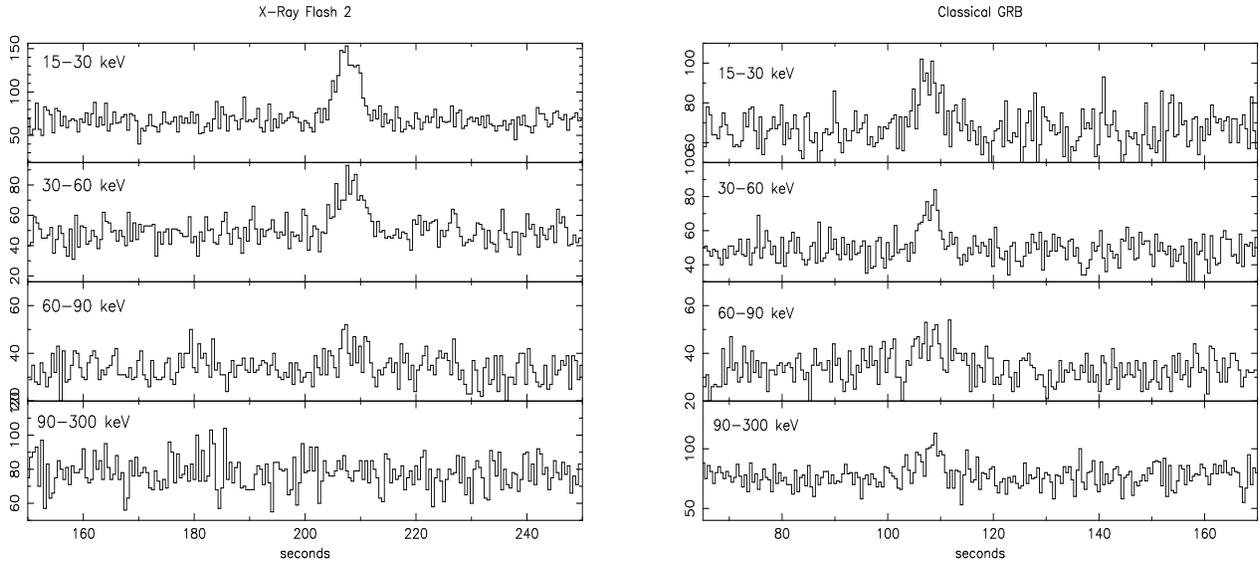

\hspace*{-1.6cm}
\begin{minipage}{3.5in}
\hfill
\epsfig{figure=h2lc.ps,height=3.0in,width=2.9in,angle=-90}
\end{minipage}
\begin{minipage}{3.5in}
\hfill
\epsfig{figure=band.ps,height=3.0in,width=2.9in,angle=-90}
\end{minipage}
\caption{Simulated light curves of an X--Ray Flash and a  {\it classical} GRB in various energy bands.}
\label{lcs}
\end{figure}

A potentially new class of transients, the X--ray Flashes \cite{heise}, has been
observed with the {\it Beppo}SAX Wide Field Cameras (2-25 keV).
These sources have spectral (non-thermal spectra) and temporal
characteristics similar to the GRBs, but they are not detected in the {\it Beppo}SAX
GRB Monitor (40-700 keV). They could be a population of
super-soft GRBs or a completely different class of sources. Some of these X--ray Flashes
(about 10) have been detected in an off-line scan of BATSE data \cite{kippen},
but they were too faint to activate the on-board trigger.
Since ISGRI has a lower energy threshold than BATSE, it will be particularly sensitive
to these events characterized by a softer spectrum.\\
We have performed some simulations using the parameters of the X--ray Flashes detected
by both the WFCs and BATSE off-line,
extrapolating the WFCs spectra (single power laws) to the ISGRI energy range.
For comparison we have simulated also a {\it classical} GRB
($\alpha$=1, $\beta$=2.25, E$_{p}$=250 keV) \cite{preece}
with the same peak flux. 
The light curves (time bin=0.5 s) in various energy bands
of one of the X--ray Flashes ($\alpha$=2.2) and of the GRB are shown in
figure \ref{lcs}.
The spectral difference between the two events is clearly visible by
comparing their light curves at different energies.


\end{document}